# A Preliminary Study for the development of an Early Method for the Measurement in Function Points of a Software Product

Ramón Asensio Monge, Francisco Sanchis Marco, Fernando Torre Cervigón ,Víctor García García, Gustavo Uría Paino

**Abstract.** The Function Points Analysis (FPA) of A.J. Albrecht is a method to determine the functional size of software products. The International Function Point Users Group, (IFPUG), establishes the FPA like a standard in the software functional size measurement. The IFPUG [3] [4] method follows the Albrecht's method and incorporates in its succesive versions modifications to the rules and hints with the intention of improving it [7]. The required documentation level to apply the method is the functional specification which corresponds to level I in the Rudolph's clasification [8]. This documentation is avalaible with some difficulty for those companies which are dedicated to develop software for third parties when they have to prepare the appropiate budget for this development. Then, we face the need of developing an early method [6] [9] for measuring the functional size of a software product that we will name to abbreviate it Early Method or EFPM (Early Function Point Method). The required documentation to apply the EFPM would be the User Requirements or some analogous documentations. This is a part of a research work now in process in Oviedo University. In this article we only show the following, results:

- From the measurements of a set of projects using the IFPUG method v. 4.1 we obtain the linear correlation coefficients between the total number of Function Points for each project and the counters of the ILFs number, ILFs+EIFs number and EIs+EOs+EQs number.
- Using the preliminary results we compute the regression functions.

This results w´ill allow us to determine the factors to be considered in the development of EFPM and to estimate the function points.

## 1 BRIEF INTRODUCTION TO THE IFPUG METHOD V 4.1 (FP4.1)

In January of 1999 the IFPUG method version 4.1 [3] [4] is publicated, which incorporates new rules, corrects others for resolving no documented situations and it is added new hints and examples which help to understand the method.
The method considers five types of function:
- Internal logical file, ILF, is a group of logical related data or control information, identifiable by the user and maintained within the application boundary.
- External interface file, EIF, is a group of logical related information or control information, identifiable by the user and referenced by the application, but maintained within other application boundary.
- External input, EI, is an elementary process or control information that crosses the application boundary from outside to inside. The primary intent of an EI is to maintain one or more ILFs and/or to alter the behaviour of the system.
- External output, EO, is an elementary process that sends data or control information outside the application boundary. The primary intent of an external output is to present information to the user through processing logic other than, or in addition to, the retrieval of data or control information. The processing logic must contain at least one mathematical formula or calculation or create derived data. One external output may also maintain one or more ILFSs and/or alter the behaviour of the system.
- External inquery, EQ, is an elementary process that sends data or control information outside the application boundary. The primary intent of an external inquery is to present information to the user through the retrieval of data or control information from ILFs and EIFs. The processing logic cannot contain mathematical formulas, calculations or create derived data. No ILFs is maintaned, nor is the behaviour of the system altered.

The complexity level for ILFs and EIFs is based on the number of RETs and DETs. A data element type, DET, is an unique user recognizable, non-repeated field. A record element type, RET, is a user recognizable subgroup of data elements within an ILF or EIF.

The complexity is determined by means of a table, whose values allow to assign the appropiate weights to each data function types according to its complexity. See table 1.

Table 1, Functional complexity of ILFs and EIFs

|                | 1 to 19 DETs | 20 to 50 DETs | 51 or more DETs |
|----------------|--------------|---------------|-----------------|
| 1 RET          | Low          | Low           | Average         |
| 2 to 5 RETs    | Low          | Average       | High            |
| 6 or more RETs | Average      | High          | High            |

The complexity level for EIs, EOs and EQs appears according to the number of FTRs and DETs. A file type referenced, FTR, is an ILF read or maintained by a transactional function or an EIF read by a transactional function. A data element type, DET, is a unique user recognizable, non-repeated field.

The complexity is determined by means of a table, whose values allow to assign the appropiate weigths for each transactional function type according to its complexity. See tables 2 and 3.

Table 2, Functional complexity of EIs

|  | 1 to 4 DETs | 5 to 15 DETs | 16 or more DETs |
|---|---|---|---|
| 0 or 1 FTR | Low | Low | Average |
| 2 FTRs | Low | Average | High |
| 3 or more FTRs | Average | High | High |

Table 3, Functional complexity of EOs and EQs

|  | 1 to 5 DETs | 6 to 19 DETs | 20 or more DETs |
|---|---|---|---|
| 0 or 1 FTR | Low | Low | Average |
| 2 or 3 FTRs | Low | Average | High |
| 4 or more FTRs | Average | High | High |

Translating the complexity level to the table 4, that we show next, it is determined the number of function points that corresponds to each function type.

Table 4, Number of Function points for each function type according to its complexity

| Function type | Complexity level | | |
|---|---|---|---|
|  | Low | Average | High |
| ILF | 7 | 10 | 15 |
| EIF | 5 | 7 | 10 |
| EI | 3 | 4 | 6 |
| EO | 4 | 5 | 7 |
| EQ | 3 | 4 | 6 |

## 2 RESULTS OF THE MEASUREMENTS

With this objective, it was necessary to measure a set of projects [1] [2] , so 30 were selected, all of them inside the management information systems that constitute the own domain of the IFPUG method v 4.1. These projects belong to organizations that include the areas of Administration, Finances, Services or Industry.

The level of documentation used was the specifications generated in the analysis phase that corresponds to the level I in the Rudolph's clasification [8], irrespective of the development phase in which the project was and still in the case it was running.

The number of raters was 4.

Each project was measured twice by different raters and the distribution of assignated projects to each rater and organization is in table 5.

Table 5, Projects for each rater and organization

| RATER | ORGANIZATION / PROJECT | | | | | | | | | | | | | | |
|---|---|---|---|---|---|---|---|---|---|---|---|---|---|---|---|
|  | ORGAN. 1 | | | ORGAN. 2 | | | ORGAN. 3 | | | ORGAN. 4 | | | ORGAN. 5 | | |
| 1 | 1 | 2 | 3 | 7 | 8 | 9 | 13 | 14 | 15 | 19 | 20 | 21 | 25 | 26 | 27 |
| 2 | 1 | 4 | 5 | 7 | 10 | 11 | 13 | 16 | 17 | 19 | 22 | 23 | 25 | 28 | 29 |
| 3 | 2 | 4 | 6 | 8 | 10 |  | 12 | 14 | 16 | 18 | 20 | 22 | 24 | 26 | 28 | 30 |
| 4 | 3 | 5 | 6 | 9 | 11 |  | 12 | 15 | 17 | 18 | 21 | 23 | 24 | 27 | 29 | 30 |

The results of the 60 measurements in number of unadjusted are showed in table 6. The values of the counters that corresponds to the number of ILFs, ILFs+EIFs and EIs+EOs+EQs are also included in this table. They are named CILF, CILFEIF and CEIEOEQ each one.

Table 6, Results of the measurements

| P | FP | CILF | CILFEIF | CEIEOEQ | P | FP | CILF | CILFEIF | CEIEOEQ |
|---|---|---|---|---|---|---|---|---|---|
| 1 | 203,0 | 8 | 8,00 | 32,00 | 16 | 247,0 | 5 | 15,00 | 33,00 |
| 1 | 379,0 | 10 | 37,00 | 31,00 | 16 | 265,0 | 5 | 12,00 | 43,00 |
| 2 | 266,0 | 8 | 11,00 | 36,00 | 17 | 370,0 | 19 | 19,00 | 54,00 |
| 2 | 284,0 | 8 | 13,00 | 46,00 | 17 | 335,0 | 17 | 21,00 | 45,00 |
| 3 | 175,0 | 2 | 7,00 | 27,00 | 18 | 438,0 | 11 | 24,00 | 71,00 |
| 3 | 171,0 | 2 | 10,00 | 22,00 | 18 | 445,0 | 12 | 23,00 | 65,00 |
| 4 | 218,0 | 5 | 15,00 | 24,00 | 19 | 349,0 | 13 | 18,00 | 50,00 |
| 4 | 218,0 | 5 | 11,00 | 36,00 | 19 | 341,0 | 13 | 17,00 | 55,00 |
| 5 | 160,0 | 0 | 23,00 | 7,00 | 20 | 256,0 | 10 | 15,00 | 32,00 |
| 5 | 119,0 | 0 | 14,00 | 8,00 | 20 | 281,0 | 10 | 15,00 | 42,00 |
| 6 | 219,0 | 10 | 14,00 | 38,00 | 21 | 127,0 | 1 | 9,00 | 14,00 |
| 6 | 240,0 | 10 | 14,00 | 38,00 | 21 | 94,0 | 1 | 7,00 | 10,00 |
| 7 | 236,0 | 9 | 10,00 | 29,00 | 22 | 118,0 | 3 | 9,00 | 16,00 |
| 7 | 268,0 | 10 | 11,00 | 37,00 | 22 | 152,0 | 5 | 11,00 | 17,00 |
| 8 | 402,0 | 15 | 20,00 | 48,00 | 23 | 244,0 | 7 | 18,00 | 26,00 |
| 8 | 346,0 | 15 | 18,00 | 47,00 | 23 | 268,0 | 7 | 19,00 | 24,00 |
| 9 | 216,0 | 2 | 16,00 | 20,00 | 24 | 208,0 | 9 | 15,00 | 23,00 |
| 9 | 227,0 | 2 | 16,00 | 20,00 | 24 | 166,0 | 5 | 10,00 | 18,00 |
| 10 | 298,0 | 16 | 20,00 | 30,00 | 25 | 258,0 | 13 | 13,00 | 42,00 |
| 10 | 246,0 | 9 | 15,00 | 27,00 | 25 | 269,0 | 13 | 13,00 | 44,00 |
| 11 | 221,0 | 4 | 7,00 | 33,00 | 26 | 403,0 | 9 | 17,00 | 53,00 |
| 11 | 155,0 | 4 | 7,00 | 17,00 | 26 | 414,0 | 9 | 17,00 | 54,00 |
| 12 | 385,0 | 20 | 22,00 | 60,00 | 27 | 609,0 | 34 | 43,00 | 84,00 |
| 12 | 487,0 | 16 | 22,00 | 67,00 | 27 | 719,0 | 34 | 47,00 | 88,00 |
| 13 | 262,0 | 9 | 12,00 | 38,00 | 28 | 277,0 | 17 | 21,00 | 34,00 |
| 13 | 292,0 | 10 | 13,00 | 47,00 | 28 | 235,0 | 15 | 17,00 | 29,00 |
| 14 | 441,0 | 15 | 26,00 | 51,00 | 29 | 120,0 | 3 | 8,00 | 15,00 |
| 14 | 462,0 | 15 | 22,00 | 67,00 | 29 | 113,0 | 3 | 7,00 | 16,00 |
| 15 | 519,0 | 16 | 26,00 | 78,00 | 30 | 234,0 | 10 | 24,00 | 21,00 |
| 15 | 577,0 | 18 | 30,00 | 65,00 | 30 | 250,0 | 10 | 25,00 | 22,00 |

## 3 LINEAR CORRELATION COEFFICIENTS AND REGRESSION FUNCTIONS

The linear correlation coefficients between the total number of function points and the counters of the number of ILFs, EIFs and EIs+EOs+EQs are showed next:

**Calculation of the correlation coefficient between the function points and the counter of ILFs**

Summary of the model

| Model | R | R squared | R squared corrected | Typical error of the estimation |
|---|---|---|---|---|
| 1 | ,848 | ,718 | ,713 | 69,0822 |

a Prediction variables: (Constant), CILF

The linear correlation coefficient has a value of 0,848 which allows us to conclude that there is a high correlation between the Function Points (FP) and the counter of ILFs (CILF).

**Calculation of the regression function between the function points and the counter of ILFs**

Coefficients

| Model | | No standard coefficients | | Standard coefficients | t | Sig. |
|---|---|---|---|---|---|---|
| | | B | Typical error | Beta | | |
| 1 | (Constant) | 130,327 | 15,755 | | 8,272 | ,000 |
| | CILF | 15,902 | 1,307 | ,848 | 12,162 | ,000 |

a Dependent variable: FP

The regression function is:

$$FP = 130{,}327 + 15{,}902 * CILF$$

Figure 1 represents the graphic of this line and as it can be observed the set of points fits very well this line. This will allow us with high confidence level to estimate the size in function points once the counter of ILFs is known.

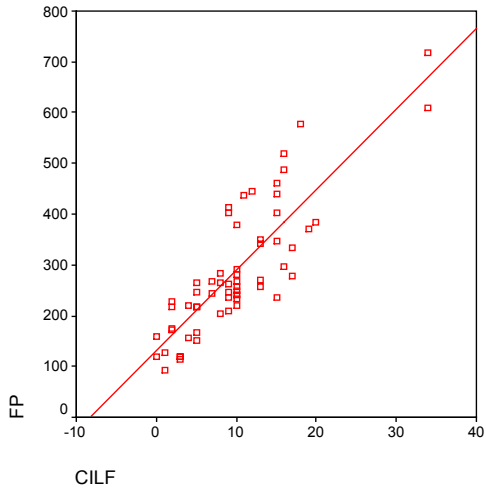

Fig. 1.- Regression line FP - CILF

**Calculation of the correlation coefficient between the function points and the counter of ILFs + EIFs**

| | | Summary of the model | | |
|---|---|---|---|---|
| Model | R | R squared | R squared corrected | Typical error of estimation |
| 1 | ,824 | ,679 | ,673 | 73,7912 |

a Prediction variables: (Constant), CILFEIF

Like in the previous case it is observed that the linear correlation coefficient has a value of 0,824 which allows us to conclude that there is a very high correlation between the Function Points (FP) and the counter of ILFs+EIFs (CILFEIF).

**Calculation of the regresión function between the function points and the counter of ILFs + EIFs**

Coefficients

| | | No standard coefficients | | Standard coefficients | t | Sig. |
|---|---|---|---|---|---|---|
| Model | | B | Typical error | Beta | | |
| 1 | (Constant) | 66,905 | 22,156 | | 3,020 | ,004 |
| | CILFEIF | 13,035 | 1,178 | ,824 | 11,067 | ,000 |

a Dependent variable: FP

The regression function is:

$$FP = 66{,}905 + 13{,}035 * CILFEIF$$

Figure 2 represents the graphic of this line and as it can be observed the set of points fits very well this line. This will allow us with high confidence level to estimate the size in function points once the counter of ILFs+EIFs is known.

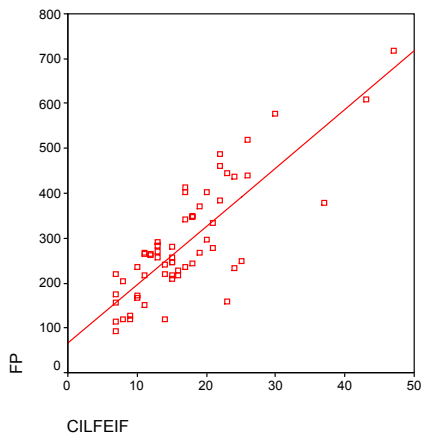

Fig. 2.- Regression line FP – CILFEIF

**Calculation of the correlation coefficient between the fnction points and the counter of EIs + EOs + EQs**

Summary of the model

| Model | R | R squared | R squared corrected | Typical error of the estimation |
|---|---|---|---|---|
| 1 | ,932 | ,869 | ,867 | 47,0237 |

a  Prediction variables: (Constant), CEIEOEQ

In this case it is observed that the linear correlation coefficient has a value of 0,932 which represents that there is a very high correlation between the Function Points (FP) and the counter of EIs+EOs+EQs (CEIEOEQ).

Coefficients

| Model | | No standard coefficients | | Standard coefficients | t | Sig. |
|---|---|---|---|---|---|---|
| | | B | Typical error | Beta | | |
| 1 | (Constant) | 50,784 | 13,521 | | 3,756 | ,000 |
| | CEIEOEQ | 6,289 | ,320 | ,932 | 19,658 | ,000 |

a  Dependent variable: FP

The regression function is:

$$FP = 50{,}784 + 6{,}289 * CEIEOEQ$$

Figure 3 represents the graphic of this line and as it can be observed the set of points fits very well this line. This will allow us with high confidence level to estimate the size in function points once the counter of EIs+EOs+Eqs is known.

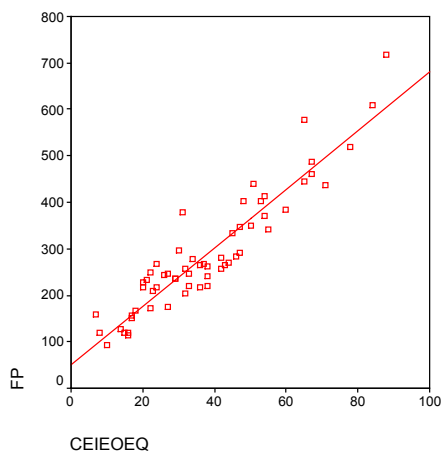

Fig. 3.- Regression line FP -CEIEOEQ

## 4 CONCLUSIONS

The strong correlation found between the number of Function Points and the counters of ILFs, ILFs+EIFs and EIs+EOs+EQs will allow us to develop an estimation method in the category of derived methods, also known as algorithmic models based methods.

The knowledge of any of the following components:
- o Logical data groups created and maintained by the application (typically the number of ILFs)
- o Logical data groups processed by the application (typically ILFs+EIFs)
- o Number of processes ( tipically EIs+EOs+EQs)

yields the factors to be used in the EFPM method.

The computed regression functions will allow us to estimate the number of Function Points of different Projects from those factors.

To complete this preliminary study, we need determine if it´s possible identify these factors using a documentation before obtaining the Functional Specifications.

The great number of projects used in the sample of these work supposes an advantage in comparison whith other derived estimation methods (Tichenor ILF Model [10] or FP Prognosis of CNV AG) wich use a very small samples [5].

**This work is part of a research project financed by the FICYT with reference number  PC-TIC01-01**